\documentclass[aoas,MSNbibl,nameyear,seceqn,dvips]{arximspdf}
\usepackage{stfloats}
\usepackage{graphicx}

\doi{10.1214/12-AOAS614} 
\volume{7}
\issue{4}
\pubyear{2013}
\firstpage{1838}
\lastpage{1865}

\makeatletter

\fnbelowfloat

\newcommand{\rrvert}{\vert}
\newcommand{\llvert}{\vert}

\newcommand{\e}{\mathrm{e}}
\renewcommand{\d}{\mathrm{d}}
\newcommand{\xmin}{x_{\min}}

\makeatother

\begin{document}
\begin{frontmatter}

\title{Estimating the historical and future probabilities of large
terrorist events\thanksref{T1}}
\runtitle{Estimating the probability of large terrorist events}

\relateddoi{T1}{Discussed in \doi{10.1214/13-AOAS614A},
\doi{10.1214/13-AOAS614B},
\doi{10.1214/13-AOAS614C},
\doi{10.1214/13-AOAS614D},
\doi{10.1214/13-AOAS614E},
\doi{10.1214/13-AOAS614F};
rejoinder at \doi{10.1214/13-AOAS614R}.}

\begin{aug}
\author[A]{\fnms{Aaron} \snm{Clauset}\corref{}\ead[label=e1]{aaron.clauset@colorado.edu}}
\and
\author[B]{\fnms{Ryan} \snm{Woodard}\ead[label=e2]{rwoodard@ethz.ch}}
\runauthor{A. Clauset and R. Woodard}
\affiliation{University of Colorado, Boulder and ETH Z\"urich}
\address[A]{Department of Computer Science and\\
\quad the BioFrontiers Institute \\
University of Colorado, Boulder\\
Boulder, Colorado 80309\\
USA\\
\printead{e1}}
\address[B]{Department of Management, Technology \\
\quad and Economics \\
ETH Z\"urich\\
Kreuzplatz 5, CH-8032 Z\"urich \\
Switzerland\\
\printead{e2}} 
\end{aug}

\received{\smonth{4} \syear{2012}}
\revised{\smonth{7} \syear{2012}}

%
\begin{abstract}
Quantities with right-skewed distributions are ubiquitous in complex
social systems, including political conflict, economics and social
networks, and these systems sometimes produce extremely large events.
For instance, the 9/11 terrorist events produced nearly 3000
fatalities, nearly six times more than the next largest event. But, was
this enormous loss of life statistically unlikely given modern
terrorism's historical record? Accurately estimating the probability of
such an event is complicated by the large fluctuations in the empirical
distribution's upper tail. We present a generic statistical algorithm
for making such estimates, which combines semi-parametric models of
tail behavior and a nonparametric bootstrap. Applied to a global
database of terrorist events, we estimate the worldwide historical
probability of observing at least one 9/11-sized or larger event since
1968 to be 11--35\%. These results are robust to conditioning on global
variations in economic development, domestic versus international
events, the type of weapon used and a truncated history that stops at
1998. We then use this procedure to make a data-driven statistical
forecast of at least one similar event over the next decade.
\end{abstract}

%

\end{frontmatter}

\section{Introduction}
The September 11th terrorist attacks were the largest such events in
modern history, killing nearly 3000 people [\citet
{mipt2008,gtd2011b}]. Given their severity, should these attacks be
considered statistically unlikely or even outliers? What is the
likelihood of another September 11th-sized or larger terrorist event,
worldwide, over the next decade?

Accurate answers to such questions would shed new light both on the
global trends and risks of terrorism and on the global social and
political processes that generate these rare events [\citet
{sornette2009,sornette2012,mcmorrow2009}], which depends in part on
determining whether the same processes generate both rare, large events
and smaller, more common events. Insights would also provide objective
guidance for our long-term expectations in planning, response and
insurance efforts [\citet{dehannferreria2006,reissthomas2007}], and
for estimating the likelihood of even larger events, including
mass-casualty chemical, biological, radioactive or nuclear (CBRN)
events [\citet{cameron2000,richardsongordon2007}].

The rarity of events like 9/11 poses two technical problems: (i) we
typically lack quantitative mechanism-based models with demonstrated
predictive power at the global scale (which is particularly problematic
for CBRN events) and (ii)~the global historical record contains few
large events from which to estimate mechanism-agnostic statistical
models of large events alone. That is, the rarity of big events implies
large fluctuations in the distribution's upper tail, precisely where we
wish to have the most accuracy. These fluctuations can lead to poor
out-of-sample predictive power in conflict
[see \citet{becketal2000}, Bueno de Mesquita (\citeyear{demesquita2003,demesquita2011}),
\citet{kingzeng2001}, \citet{rustadetal2011}, \citet{wardetal2010}]
and can complicate both selecting the correct model of the tail's
structure and accurately estimating its
parameters [\citet{clausetetal2009}]. Misspecification can lead to
severe underestimates of the true probability of large events, for
example, in classical financial risk
models [\citet{farmerlillo2004,fcir2011}].

Little research on terrorism has focused on directly modeling the
number of deaths (``severity'')\setcounter{footnote}{1}\footnote{Other notions of event ``size'' or severity, which we do not
explore here, might be the economic cost, number injured, political
impact, etc. To the extent that such notions may be quantitatively
measured, our algorithm could also be applied to them.}
in individual terrorist events [\citet{mcmorrow2009}]. When deaths are
considered, they are typically aggregated and used as a covariate to
understand other aspects of terrorism, for example, trends over
time [Enders and Sandler (\citeyear{enderssandler2000,enderssandler2002})], the when,
where, what, how and why of the resort to terrorism [\citet
{browndaltonhoyle2004,enderssandler2006,valenzuelaetal2010}],
differences between organizations [\citet{asalrethemeyer2008}], or
the incident rates or outcomes of events [\citet
{enderssandler2000,endersetal2011}]. Such efforts have used time
series analysis [Enders and Sandler (\citeyear{enderssandler2000,enderssandler2002}), \citet{endersetal2011}],
qualitative models or human expertise of specific scenarios, actors,
targets or attacks [\citet{wulfetal2003}] or quantitative models
based on factor analysis [\citet{li2005,pape2003}], social
networks [\citet{sageman2004,desmaraiscranmer2011}] or formal
adversarial interactions [\citet
{major1993,kardeshall2005,enderssandler2006}]. Most of this work
focuses on modeling central tendencies, treats large events like 9/11
as outliers, and says little about their quantitative
probability [\citet{clausetetal2007}] or their long-term hazard.

Here, we describe a statistical algorithm for estimating the
probability of large events in complex social systems in general, and
in global terrorism in particular. Making only broad-scale and
long-term probabilistic estimates, our approach is related to
techniques used in seismology, forestry, hydrology and natural disaster
insurance to estimate the probabilities of individual rare catastrophic
events [\citet
{gutenbergrichter1944,gumbel1941,reedmckelvey2002,breimanstonekooperberg1990,dehannferreria2006,reissthomas2007}].
Our approach combines maximum-likelihood methods, multiple models of
the distribution's tail and computational techniques to account for
both parameter and model uncertainty. It provides a quantitative
estimate of the probability, with uncertainty, of a large event. The
algorithm also naturally generalizes to include certain event
covariates, which can shed additional light on the probability of large
events of different types.

Using this algorithm to analyze a database of 13,274 deadly terrorist
events worldwide from 1968--2007, we estimate the global historical
probability of at least one 9/11-sized or larger terrorist event over
this period to be roughly 11--35\%. Furthermore, we find the nontrivial
magnitude of this historical probability to be highly robust, a direct
consequence of the highly right-skewed or ``heavy-tailed'' structure of
event sizes [\citet{clausetetal2007}]. Thus, an event the size or
severity of the September 11th terrorist attacks, compared to the
global historical record, should not be considered a statistical
outlier or even statistically unlikely. Using three potential scenarios
for the evolution of global terrorism over the next decade, we then
estimate the worldwide future probability of a similarly large event as
being not significantly different from the historical level. We close
by discussing the implications for forecasting large terrorist events
in particular and for complex social systems in general.

\section{Estimating the probability of a large event}
The problem of estimating the probability of some observed large event
is a kind of tail-fitting problem, in which we estimate parameters for
a distributional model using only the several largest observations.
This task is distinct from estimating the distribution of maxima within
a sample [\citet{dehannferreria2006,reissthomas2007}], and is more
closely related to the peaks-over-threshold literature in hydrology,
seismology, forestry, finance and insurance [\citet
{adleretal1998,breimanstonekooperberg1990,dehannferreria2006,gumbel1941,gutenbergrichter1944,reedmckelvey2002,reissthomas2007,resnick2006}].
Here, we aim specifically to deal with several sources of uncertainty
in this task: uncertainty in the location of the tail, uncertainty in
the tail's true structure, and uncertainty in the model
parameters.\vadjust{\goodbreak}

Our approach is based on three key insights. First, because we are
interested only in rare large events, we need only model the structure
of the distribution's right or upper tail, which governs their
frequency. This replaces the difficult problem of modeling both the
distribution's body and tail [\citet
{resnick2006,dehannferreria2006,reissthomas2007}] with the less
difficult problem of identifying a value $\xmin$ above which a model
of the tail alone fits well.\footnote{The notation $\xmin$ should not be confused with the first
order statistic, $x_{(1)}=\min_{i}x_{i}$.}
That is, choose some $\xmin$ and a tail model $\Pr(x | \theta,\xmin)$
defined on $x\in[\xmin,\infty)$. We will revisit the problem of
choosing $\xmin$ below.

Second, in complex social systems, the correct tail model is typically
unknown and a poor choice may lead to severe misestimates of the true
probability of a large event. We control for this model uncertainty by
considering multiple tail models. Given these models and a common
choice of $x_{\min}$, we use a likelihood ratio test to identify and
discard the statistically implausible ones [\citet{clausetetal2009}].
In principle, the remaining models could be averaged to produce a
single estimate with confidence intervals [\citet
{claeskenshjort2008}], for example, to aid decision makers. We return
to this point in more detail below.

Finally, large fluctuations in the distribution's upper tail occur
precisely where we wish to have the most accuracy, leading to parameter
uncertainty. Using a nonparametric bootstrap [\citet
{efrontibshirani1993}] to simulate the generative process of event
sizes, we incorporate the empirical data's inherent variability into
the estimated parameters, weight models by their likelihood under the
bootstrap distribution and construct extreme value confidence
intervals [\citet{breimanstonekooperberg1990}].

This combination of techniques provides a statistically principled and
data-driven solution for estimating the probability of observing rare
events in empirical data with unknown tail structure. If such an event
is observed, the algorithm provides a measure of whether its occurrence
was in fact unlikely, given the overall structure of the distribution's
tail. For instance, if the estimated probability is negligible (say,
$p<0.01$), the event may be judged statistically unlikely. When several
tail models are plausible and agree that the probability is away from
$p=0$, the event can be judged to be statistically likely, despite the
remaining uncertainty in the tail's structure.

\subsection{The method}
Our goal is to estimate the probability that we would observe at least
$\ell$ ``catastrophic'' events of size $x$ or greater in an empirical
sample.\footnote{Consider events to be generated by a kind of marked point
process [\citet{lastbrandt1995}], where marks indicate either the
event's severity or that it exceeded some threshold~$x$. Although we
assume the number of marks $n$ to be fixed, this could be relaxed to
incorporate additional uncertainty into the algorithm's output.}
In principle, any size $x$ and any value $\ell$ may be chosen, but, in
practice, we typically choose $x$ as the largest (and thus rarest)
event in the empirical data and set $\ell=1$. To ensure that our
estimate is meaningful from a historical perspective, we remove the
catastrophic event(s) from the empirical sample before applying the
algorithm. Here we describe the method in terms of univariate
distributions, but its generalization to certain covariates is
straightforward (see Appendix \ref{appendixweapons}).

Let $\Pr(x | \theta,\xmin)$ denote a particular tail model with
parameters $\theta$, let $\{x_{i}\}$ denote the $n$ empirical event
sizes (sans the catastrophic events), and let $Y=\{y_{j}\}$ be a
bootstrap of these data ($n$ samples drawn from $\{x_{i}\}$ with
replacement). To begin, we assume a fixed $\xmin$, the smallest value
for which the tail model holds, and later describe the generalization
to variable $\xmin$.

The fraction of empirical events with values in the tail region is
$p_{\mathrm{tail}} = \#\{x_{i}\geq\xmin\}/n$, and in each bootstrap the
number is a binomial random variable with probability $p_{\mathrm{tail}}$:
%
\begin{equation}\label{eqntail}
n_{\mathrm{tail}}\sim\operatorname{Binomial}(n,p_{\mathrm{tail}}).
\end{equation}
The maximum likelihood estimate $\hat{\theta}$ is a deterministic
function of the portion of $Y$ above $\xmin$, which we denote $\theta
(Y,\xmin)$.

Given that choice, the probability under the fitted model that not one
of $n'_{\mathrm{tail}}=1+n_{\mathrm{tail}}$ events is at least as big as $x$ is
%
\begin{equation}
\label{eqnoevents} F \bigl(x | \theta(Y,\xmin) \bigr)^{n'_{\mathrm{tail}}} =
\biggl(
\int_{x_{\min}}^{x} \Pr(y | \hat{\alpha},\xmin) \,\d y
\biggr)^{n'_{\mathrm{tail}}}.
\end{equation}
Thus, $1-F (x | \theta(Y,\xmin) )^{n'_{\mathrm{tail}}}$ is
the probability that at least one event is of catastrophic size.
Because the bootstrap $Y$ is itself a random variable, to derive the
marginal probability of observing at least one catastrophic event, we
must integrate the conditional probability over the domain of the
bootstrap distribution:
%
\begin{eqnarray}\label{eqbootstrappdf}\qquad
p(n_{\mathrm{tail}},\theta) & = & p(n_{\mathrm{tail}},Y)
\nonumber\\[-8pt]\\[-8pt]
& = & \int{\d y_1 \cdots\d y_{n_{\mathrm{tail}}} \bigl(1 - F\bigl(x;\theta
(Y,\xmin)\bigr)^{n'_{\mathrm{tail}}} \bigr) \prod_{i=1}^{n_{\mathrm{tail}}}{r(y_{i}
| n_{\mathrm{tail}}).}}\nonumber
\end{eqnarray}
The trailing product series here is the probability of drawing the
specific sequence of values $y_{1}, \ldots, y_{n_{\mathrm{tail}}}$ from the
fixed bootstrap distribution $r$. Finally, the total probability $p$ of
at least one catastrophic event is given by a binomial sum over this
equation.\footnote{We may calculate $p$ in either of two ways: (i) we draw
$n_{\mathrm{tail}}$ events from a tail model alone, or (ii) we draw $n$
events from a conditional model, in which the per-event probability is
$q(x) = \Pr(X\geq x | X\geq x_{\mathrm{min}}) \Pr(X\geq x_{\min
})=p_{\mathrm{tail}}(1-F(x | \theta,x_{\min}))$. When the probability of
a catastrophic event is small, these calculations yield equivalent results.}

When the correct value $\xmin$ is not known, it must be estimated
jointly with $\theta$ on each bootstrap. Maximum likelihood cannot be
used for this task, because $\xmin$ truncates $Y$. Several principled
methods for automatically choosing $\xmin$ exist, for example,
\citet
{breimanstonekooperberg1990,clausetetal2009,clausetetal2007,danielssonetal2001,dekkersdehann1993,dreeskaufmann1998,hancockjones2004}.
So long as the choice of $\xmin$ is also a deterministic function of
$Y$, the above expression still holds. Variation in $\xmin$ across the
bootstraps, however, leads to different numbers of observations $n_{\mathrm{tail}}$ in the tail region. The binomial probability $p_{\mathrm{tail}}$ is
then itself a random variable determined by $Y$, and $n_{\mathrm{tail}}$ is
a random variable drawn from a mixture of these binomial distributions.

Analytically completing the above calculation can be difficult, even
for simple tail models, but it is straightforward to estimate
numerically via Monte Carlo:
\begin{enumerate}
\item\label{step1} Given $n$ empirical sizes, generate $Y$ by
drawing $y_{j}$, $j=1,\ldots,n$, uniformly at random, with replacement,
from the observed $\{x_{i}\}$ (sans the $\ell$ catastrophic events).
\item\label{step2} Jointly estimate the tail model's parameters
$\theta$ and $x_{\mathrm{min}}$ on $Y$, and compute $n_{\mathrm{tail}} = \#\{
y_{j}\geq\hat{x}_{\min}\}$ (see Appendix \ref{appendixmethod}).
\item\label{step3} Set $\rho=1-F(x;\hat{\theta})^{\ell+n_{\mathrm{tail}}}$, the probability of observing at least $\ell$ catastrophic
events under this bootstrap model.
\end{enumerate}
Averaging over the bootstraps yields the estimated probability $\hat
{p}=\langle\rho\rangle$ of observing at least $\ell$
catastrophic-sized events. The convergence of $\hat{p}$ is guaranteed
so long as the number of bootstraps (step \ref{step1}) tends to
infinity [\citet{efrontibshirani1993}]. Confidence intervals on
$\hat
{p}$ [\citet{breimanstonekooperberg1990,efrontibshirani1993}] may
be constructed from the distribution of the $\rho$ values. If the tail
model's c.d.f. $F(x;\theta)$ in step \ref{step3} cannot be computed
analytically, it can often be constructed numerically; failing that,
$\rho$ may always be estimated by sampling directly from the fitted model.

\subsection{Model comparison and model averaging}
In complex social systems, we typically do not know a priori
which particular tail model is correct, and the algorithm described
above will give no warning of a bad choice [but see \citet
{clausetetal2009}]. This issue is partly mitigated by estimating
$\xmin$, which allows us to focus our modeling efforts on the upper
tail alone. But, without additional evidence of the model's statistical
plausibility, the estimate $\hat{p}$ should be treated as provisional.

Comparing the results from multiple tail models provides a test of
robustness against model misspecification, for example, agreement
across models that $\hat{p}>0.01$ strengthens the conclusion that the
event is not statistically unlikely. However, wide confidence intervals
and disagreements on the precise probability of a large event reflect
the inherent difficulty of identifying the correct tail structure.

To select reasonable models to compare, standard model comparison
approaches may be used, for example, a fully Bayesian approach
[\citet
{kassraftery1995}], cross-validation [\citet{stone1974}] or minimum
description length [\citet{grunwald2007}]. Here, we use a
goodness-of-fit test to establish the plausibility of the power-law
distribution [\citet{clausetetal2009}] and Vuong's likelihood ratio
test [\citet{clausetetal2009,vuong1989}] to compare it with
alternatives. This approach has the advantage that it can fail to
choose one model over another if the difference in their likelihoods is
statistically insignificant, given the data.

In some circumstances, we may wish to average the resulting models to
produce a single estimate with confidence intervals, for example, to
aid decision makers. However, averaging poses special risks and
technical problems for estimating the probability of large events. For
instance, traditional approaches to averaging can obscure the inherent
uncertainty in the tail's structure and can produce spuriously precise
confidence intervals [\citet
{claeskenshjort2008,hjortclaeskens2003}]; a Bayesian approach would
be inconsistent with our existing framework; and an appropriate
frequentist framework is not currently available, although one may be
possible using insights from \citet{cesa-bianchietal2004}.

Thus, in our application below, we elect not to average and instead we
present results for each model. Even without averaging, however,
several valuable insights may be drawn.

\subsection{Tests of the method's accuracy} To test the accuracy of
our estimation algorithm, we examine its ability to recover the true
probability of a rare event from synthetic data with known structure.
To generate these synthetic data, we use the power-law distribution
%
\begin{equation}
\Pr(y) \propto y^{-\alpha},
\end{equation}
where $\alpha>1$ is the ``scaling'' parameter and $y\geq\xmin>0$.
When $\alpha<2$, this distribution exhibits infinite variance and
produces extreme fluctuations in the upper tail of finite-size samples.
By defining a catastrophic event $x$ to be the largest generated event
within the $n$ synthetic values, we make the test particularly
challenging because the largest value exhibits the greatest
fluctuations of all. Detailed results are given in Appendix \ref
{appendixaccuracy}.

We find that despite the large fluctuations generated by the power-law
distribution, the algorithm performs well: the mean absolute error
$\langle| \hat{p} - p | \rangle$ is small even for samples with less
than 100 events, and decays like $O(n^{-1/3})$. A~small absolute
deviation, however, may be an enormous relative deviation, for example,
if the true probability tends to zero or one. Our algorithm does not
make this type of error: the mean ratio of the estimated and true
probabilities $\langle\hat{p} / p \rangle$ remains close to 1 and
thus the estimate is close in relative terms, being only a few percent
off for $n\gtrsim100$ events.

\section{Historical probability of 9/11} Having described our
statistical approach, we now use it to estimate the historical
probability of observing worldwide at least one 9/11-sized or larger
terrorist event.

Global databases of terrorist events show that event severities (number
of deaths) are highly right-skewed or ``heavy tailed'' [\citet
{mipt2008,gtd2011b}]. We use the RAND-MIPT database [\citet
{mipt2008}], which contains 13,274 deadly events worldwide from
1968--2007. The power law is a statistically plausible model of this
distribution's tail, with $\hat{\alpha}=2.4\pm0.1$, for $x\geq\hat
{x}_{\min}=10$ [\citet{clausetetal2007,clausetetal2009}]. A
goodness-of-fit test fails to reject this model of tail event
severities ($p=0.40\pm0.03$ via Monte Carlo [\citet
{clausetetal2009}]), implying that the deviations between the
power-law model and the empirical data are indistinguishable from
sampling noise.

This fact gives us license to treat as i.i.d. random variables the
severity of these events. This treatment does force a particular and
uncommon theoretical perspective on terrorism, in which a single global
``process'' produces events, even if the actions of individual
terrorists or terrorist organizations are primarily driven by local
events. This perspective has much in common with statistical physics,
in which particular population-level patterns emerge from a sea of
individual interactions. We discuss limitations of this perspective in
Section \ref{sectionimprovements}.

Past work shows that this apparent power-law pattern in global
terrorism is remarkably robust. Although the estimated value of $\alpha
$ varies somewhat with time [\citet{clausetetal2007}], the power-law
pattern itself seems to persist over the 40-year period despite large
changes in the international system. It also appears to be independent
of the type of weapon (explosives, firearms, arson, knives, etc.), the
emergence and increasing frequency of suicide attacks, the demise of
many terrorist organizations, the economic development of the target
country [\citet{clausetetal2007}] and organizational covariates like
size (number of personnel), age and experience (total number of
attacks) [\citet{clausetgleditsch2009}].

%
\begin{figure}

\includegraphics{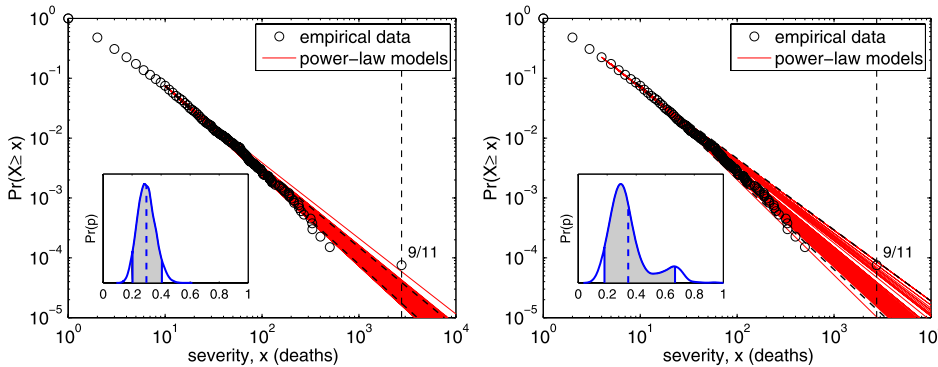}

\caption{Empirical severity distribution with 100 bootstrap power-law
models for \textup{(a)} fixed $\xmin=10$ and \textup{(b)} estimated
$\xmin$. Overprinting illustrates the ensemble of estimated models
(dashed lines show 90\% CI on $\hat{\alpha}$) and the inherent
uncertainty in the tail structure. Insets show the 90\% confidence
intervals for the estimated probability of observing at least one
9/11-sized event.}
\label{figbootstrapalpha}
\end{figure}

Comparing the power-law tail model against log-normal and stretched
exponential (Weibull) distributions, via a likelihood ratio test,
yields log-likelihood ratios of $\mathcal{R}=-0.278$ ($p=0.78$) and
$0.772$ ($p=0.44$), respectively [\citet{clausetetal2009}]. However,
neither of these values is statistically significant, as indicated by
the large $p$-values for a test against $\mathcal{R}=0$. Thus, while
the power-law model is plausible, so too are these alternatives. This
ambiguity illustrates the difficulty of correctly identifying the
tail's structure and reinforces the need to use multiple tail models in
estimating the likelihood of a rare event like 9/11. Furthermore, it
implies that slight visual deviations in the empirical distribution's
upper tail (see Figure \ref{figbootstrapalpha}) should not be
interpreted as support either for or against any of these models. In
what follows, we consider estimates derived from all three.

To apply our algorithm to this problem, we must make several choices.
For consistency with past work on the frequency of severe terrorist
events [\citet{clausetetal2007,clausetetal2009}], we choose $\xmin
$ automatically by minimizing the Kolmogorov--Smirnov good\-ness-of-fit
statistic between the tail model and the truncated empirical
data.\footnote{\citet{clausetetal2009} provide a thorough motivation of
this strategy. Briefly, the KS statistic will be large either when
$\xmin$ is too small (including nonpower-law data in the power-law
fit) or when $\xmin$ is too large (when sample size is reduced and
legitimately power-law data thrown out), but will be small between
these two cases.}
We use the discrete power-law distribution as our tail model (which
implies $\xmin$ is also discrete; see Appendix \ref{appendixmethod})
and compare its estimates to those made using log-normal and stretched
exponential models. To avoid the problem of choosing an appropriate
event count distribution, we keep the number of events $n$ fixed.

Finally, using the RAND-MIPT event data (other sources [\citet
{gtd2011b}] yield similar results; see Appendix \ref{appendixgtd}),
we define $x\geq2749$ to be a ``catastrophic'' event---the reported
size of the New York City 9/11 events.\footnote{Official sources differ slightly on the number killed in New
York City. Repeating our analyses with other reported values does not
significantly change our estimates.}
Removing this event from the empirical data leaves the largest event as
the 14 August 2007 coordinated truck bombing in Sinjar, Iraq, which
produced approximately 500 fatalities. To illustrate the robustness of
our results, we consider estimates derived from fixed and variable
$\xmin$ and from our three tail models. We also analyze the impact of
covariates like domestic versus international, the economic development
of the target country and the type of weapon used.

\subsection{Uncertainty in the scaling parameter} Let $\xmin=10$ be
fixed. Figure \ref{figbootstrapalpha}(a) shows 100 of the
fitted bootstrap models, illustrating that by accounting for the
uncertainty in $\alpha$, we obtain an ensemble of tail models and thus
an ensemble of probability estimates for a catastrophic-sized event.
The bootstrap parameter distribution $\Pr(\hat{\alpha})$ has a mean
$\langle\hat{\alpha} \rangle= 2.40$, which agrees with the maximum
likelihood value $\hat{\alpha}=2.4$ [\citet{clausetetal2009}].

To estimate the historical probability of 9/11, we use 10,000
bootstraps with $\xmin$ fixed. Letting $p$ denote the overall
probability from the algorithm, we find $\hat{p} = 0.299$, with 90\%
confidence intervals of $[0.203,0.405]$ [Figure \ref
{figbootstrapalpha}(a) inset], or about a 30\% chance over
the 1968--2007 period.

An event that occurs with probability $0.299$ over 40 years is not a
certainty. However, for global terrorism, this value is uncomfortably
large and implies that, given the historical record, the size of 9/11
should not be considered a statistical fluke or outlier.

\subsection{Uncertainty in the tail location} A fixed choice of $\xmin
$ underestimates the uncertainty in $p$ due to the tail's unknown
structure. Jointly estimating $\alpha$ and $\xmin$ yields similar
results, but with some interesting differences. Figure \ref
{figbootstrapalpha}(b) shows 100 of the bootstrap models. The
distribution of $\hat{x}_{\min}$ is concentrated at $x_{\min}=9$ or
$10$ (48\% of samples), with an average scaling exponent of $\langle
\hat{\alpha} \rangle= 2.40$. However, 15\% of models choose $x_{\min
}=4$ or $5$, and these produce much heavier-tailed models, with
$\langle\hat{\alpha} \rangle= 2.21$.

This bimodal distribution in $\hat{\alpha}$ is caused by slight
curvature in the empirical mid-to-upper tail, which may arise from
aggregating multiple types of local events into a single global
distribution (see Appendix \ref{appendixweapons}). The algorithm,
however, accounts for this curvature by automatically estimating a
slightly wider ensemble of models, with correspondingly greater density
in the catastrophic range. As a result, the estimated probability is
larger and the confidence intervals wider. Using 10,000 bootstraps, we
find $\hat{p} = 0.347$, with 90\% confidence intervals of $[0.182,
0.669]$, or about a 35\% chance over the 1968--2007 period.

%
\begin{figure}

\includegraphics{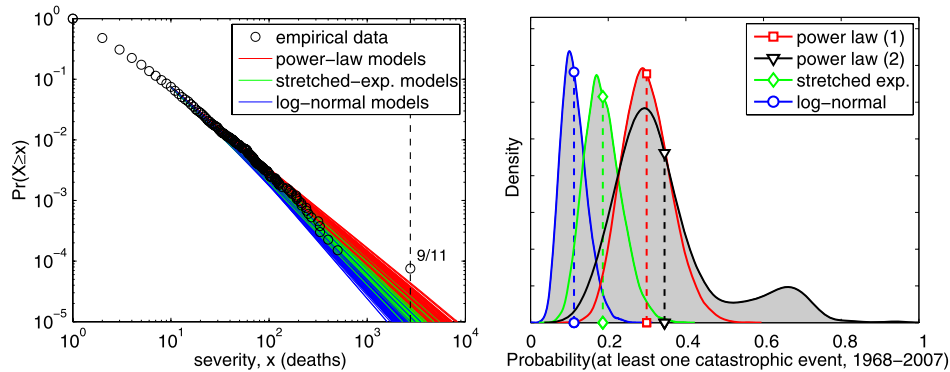}

\caption{\textup{(a)} Empirical event severities with 100 bootstrap
models for
the power-law, log-normal and stretched exponential tail models, with
$\xmin=10$ fixed. \textup{(b)} Bootstrap distributions of $\hat{p}$
for each
model, with overall estimates (Table \protect\ref{tableestimates})
given by
dashed lines.}
\label{figbootstrapalts}
\end{figure}
%

%
\begin{table}
\caption{Estimated per-event and worldwide historical probabilities
for at least one catastrophic event over the period 1968--2007, for
four tail models}
\label{tableestimates}
\begin{tabular*}{\tablewidth}{@{\extracolsep{\fill}}lcccc@{}}
\hline
& & \textbf{Est.} $\bolds{\Pr(x\geq2749)}$ &
\textbf{Est. prob.} $\bolds{p}$, & \textbf{90\% CI} \\
\textbf{Tail model} & \textbf{Parameters} & \textbf{per event,} $\bolds{q(x)}$
& \textbf{1968--2007} & \textbf{(bootstrap)} \\
\hline
Power law (1) & $\Pr
(\hat{\alpha})$, $\xmin=10$ & $0.0000270200$ & $0.299$ & $[0.203,
0.405]$ \\
Power law (2) & $\Pr(\hat{\alpha},\hat{x}_{\min})$ &
$0.0000346345$ & $0.347$ & $[0.182, 0.669]$ \\
Stretched exp. & $\Pr
(\hat{\beta},\hat{\lambda})$, $\xmin=10$ & $0.0000156780$ &
$0.187$ &$[0.115, 0.272]$ \\
log-normal & $\Pr(\hat{\mu},\hat
{\sigma})$, $\xmin=10$ & $0.0000090127$ & $0.112$ & $[0.063, 0.172]$\\
\hline
\end{tabular*}
\end{table}
%

\subsection{Alternative tail models} Comparing these estimates with
those derived using log-normal and stretched exponential tail models
provides a check on their robustness, especially if the alternative
models yield dramatically different estimates.

The mathematical forms of the alternatives are
\begin{eqnarray*}
\mbox{log-normal} \quad \Pr(x) &\propto& x^{-1} \exp\bigl[ -{(\ln x-
\mu)^2 / 2\sigma^2 } \bigr],
\\
\mbox{stretched exp.} \quad \Pr(x) &\propto& x^{\beta-1} \e^{-\lambda
x^\beta},
\end{eqnarray*}
where we restrict each to a ``tail'' domain $\xmin\leq x < \infty$
(see Appendix \ref{appendixmethod}). In the stretched exponential,
$\beta<1$ produces a heavy-tailed distribution; in the log-normal,
small values of $\mu$ and large values of $\sigma$ yield heavy tails.
Although both decay asymptotically faster than any power law, for
certain parameter choices, these models can track a power law over
finite ranges, which may yield only marginally lower estimates of large
events.\footnote{The question of power law versus nonpower law [\citet
{clausetetal2009}] is not always academic; for instance,
macroeconomic financial models have traditionally and erroneously
assumed nonpower-law tails that assign negligible probability to large
events like widespread subprime loan defaults [\citet{fcir2011}].}

To simplify the comparison between the tail models, we fix $\xmin=10$
and use 10,000 bootstraps for each fitted alternative tail model. This
yields $\hat{p}=0.112$ (CI: $[0.063, 0.172]$) for the log-normal and
$\hat{p}=0.187$ (CI: $[0.115, 0.272]$) for the stretched exponential,
or roughly an 11\% and 19\% chance, respectively. These values are
slightly lower than the estimates from the power-law model, but they
too are consistently away from $p=0$, which reinforces our conclusion
that the size of 9/11 should not be considered a statistical outlier.

Figure \ref{figbootstrapalts}(a) shows the fitted ensembles
for all three fixed-$\xmin$ tail models, and Figure \ref
{figbootstrapalts}(b) shows the bootstrap distributions $\Pr
(\hat{p})$ for these models, as well as the one with $\xmin$ free.
Although the bootstrap distributions for the log-normal and stretched
exponential are shifted to the left relative to the two power-law
models, all distributions overlap and none place significant weight
below $p=0.01$. The failure of the alternatives to disagree with the
power law can be attributed to their estimated forms roughly tracking
the power law's over the empirical data's range, which leads to similar
probabilistic estimates of a catastrophic event.

\subsection{Impact of covariates} Not all large terrorist events are
of the same type, and thus our overall estimate is a function of the
relative empirical frequency of different covariates and the structure
of their marginal distributions. Here, we apply our procedure to the
distributions associated with a few illustrative categorical event
covariates to shed some additional light on the factors associated with
large events. A generalization to and systematic analysis of arbitrary
covariates is left for future work.

For instance, international terrorist events, in which the attacker and
target are from different countries, comprise 12\% of the RAND-MIPT
database and exhibit a much heavier-tailed distribution, with $\hat
{\alpha}=1.93\pm0.04$ and $\hat{x}_{\min}=1$ (see Appendix~\ref
{appendix1998}). This heavier tail more than compensates for their
scarcity, as we estimate $\hat{p}=0.475$ (CI: $[0.309, 0.610]$;
Figure \ref{figcovariate}(a)) for at least one such
catastrophic event from 1968--2007.\footnote{The implication of a larger $\hat{p}$ for a covariate
distribution, as compared to the full data set, is a smaller $p$ for
the excluded types of events. That is, a larger $p$ for international
events implies a smaller $p$ for domestic events.}
A similar story emerges for events in economically developed nations,
which comprise 5.3\% of our data (see Appendix \ref{appendixoecd}).
Focusing on such large events ($x\geq10$), we estimate $\hat
{p}=0.225$ (CI: $[0.037, 0.499]$, Figure~\ref{figcovariate}(b)).

Another important event covariate is the type of weapon used. The tails
of the weapon-specific distributions remain well described as power
laws, but weapons like guns, knives and explosives exhibit less heavy
tails (fewer large events) than unconventional weapons [\citet
{clausetetal2007}], even as the former are significantly more common
than the latter. The estimation algorithm used above can be generalized
to handle categorical event covariates, and produces both marginal and
total probability estimates (see Appendix \ref{appendixweapons}).
Doing so yields an overall estimate of $\hat{p}=0.564$ (CI: $[0.338,
0.839]$; Figure \ref{figweapons}). Examining the marginal hazard
rates, we see that the largest contribution comes from explosives,
followed by fire and firearms.

\section{Statistical forecasts}
\label{secfutureterrorism} If the social and political processes
that generate terrorist events worldwide are roughly stationary, our
algorithm can be used to make principled statistical forecasts about
the future probability of a catastrophic event. Although here we make
the strong assumption of stationarity, this assumption could be relaxed
using nonstationary forecasting techniques [\citet
{cairesferreira2005,clementshendry1999,shalizietal2010}].

\begin{figure}

\includegraphics{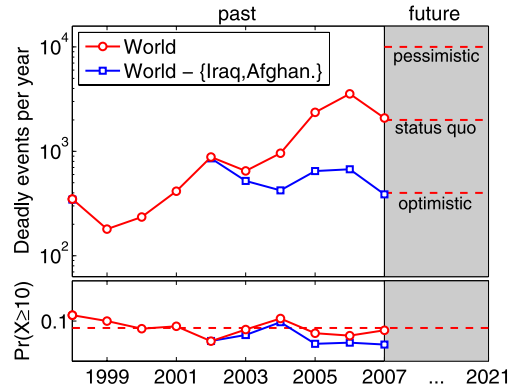}

\caption{(Upper) number of deadly (domestic and international)
terrorist events worldwide for the 10-year period 1998--2007, and
three forecast scenarios. (Lower) fraction of events that are severe,
killing at least 10 individuals and its 10-year average (dashed line).}
\label{figtimeseries}
\end{figure}

A simple forecast requires estimating the number of events $n$ expected
over the fixed forecasting horizon $t$. Using the RAND-MIPT data as a
starting point, we calculate the number of annual deadly events
worldwide $n_{\mathrm{year}}$ over the past 10 years. Figure \ref
{figtimeseries} shows the empirical trend for deadly terrorist events
worldwide from 1998--2007, illustrating a 20-fold increase in
$n_{\mathrm{year}}$, from a low of 180 in 1999 to a high of 3555 in
2006. Much of the increase is attributable to conflicts in Iraq and
Afghanistan; excluding events from these countries significantly
reduces the increase in $n_{\mathrm{year}}$, with the maxima now being 857
deadly events in 2002 and 673 in 2006. However, the fraction of events
that are severe ($x\geq10$) remains constant, averaging $\langle
p_{\mathrm{tail}}\rangle=0.082684$ (or about 8.3\%) in the former case
and $0.072601$ (or about 7.3\%) in the latter.

%
%

An estimated trend over the next decade could be obtained via fitting
standard statistical models to annual data or by soliciting judgements
from domain experts about specific conflicts. For instance, Iraq and
Afghanistan may decrease their production rates of new events over the
next decade, leading $n_{\mathrm{year}}$ to decrease unless other conflicts
replace their contributions. Rather than make potentially overly
specific predictions, we instead consider three rough scenarios (the
future's trajectory will presumably lay somewhere between): (i) an
optimistic scenario, in which the average number of terrorist attacks
worldwide per year returns to its 1998--2002 level, at about $\langle
n_{\mathrm{year}}\rangle=400$ annual events; (ii) a status quo scenario,
where it remains at the 2007 level, at about 2000 annual events; and
finally (iii) a pessimistic scenario, in which it increases to about
10,000 annual events.\footnote{Modeling these rough event counts via a Poisson process with
rate $\lambda_{\mathrm{scenario}}$ would refine our forecasts slightly.
More detailed event production models could also be used.}

A quantitative statistical forecast is then obtained by applying the
estimation algorithm to the historical data (now including the 9/11
event) and then generating synthetic data with the estimated number of
future events $n_{\mathrm{tail}}$. For each scenario, we choose
$n_{\mathrm{decade}}=10\times n_{\mathrm{year}}$ and choose
$n_{\mathrm{tail}}$ via equation (\ref{eqntail}) with
$p_{\mathrm{tail}}=0.082684$ (its historical average). Finally, we fix
$\xmin=10$ to facilitate comparison with our alternative tail models.

%
\begin{table}
\caption{Forecast estimates of at least one catastrophic event
worldwide over a 10-year period, using three tail models in each of
three forecast scenarios}
\label{tableforecasts}
\begin{tabular*}{\tablewidth}{@{\extracolsep{\fill}}lccc@{}}
\hline
& \multicolumn{3}{c@{}}{$\bolds{\Pr(x\geq2749)}$
\textbf{forecast, 2012--2021}} \\[-4pt]
& \multicolumn{3}{c@{}}{\hrulefill}\\
& \textbf{``Optimistic''} & \textbf{``Status quo''} & \textbf{``Pessimistic''} \\
\textbf{Tail model} &
$\bolds{n_{\mathrm{year}}\approx400}$ & $\bolds{n_{\mathrm{year}}\approx2000}$
& $\bolds{n_{\mathrm{year}}\approx10\mbox{\textbf{,}}000}$ \\
\hline
Power law & $0.117$ & $0.461$ & $0.944$
\\
Stretched exp. & $0.072$ & $0.306$ & $0.823$ \\
log-normal &
$0.043$ & $0.193$ & $0.643$ \\
\hline
\end{tabular*}
\end{table}
%

Table \ref{tableforecasts} summarizes the results, using 100,000
bootstraps for each of the three tail models in the three forecast
scenarios. Under the status quo scenario, all three models forecast a
19--46\% chance of at least one catastrophic event worldwide in the
next decade. In the optimistic scenario, with events worldwide being
about 5 times less common, the models forecast a 4--12\% chance. These
estimates depend strongly on the overall frequency of terrorist events
$n_{\mathrm{year}}$. Thus, the greater the popularity of terrorism
worldwide, that is, the more often terrorist attacks are launched, the
greater the general likelihood that at least one will be catastrophic.
Any progress in moving the general frequency of terrorism toward the
more optimistic scenario is likely to reduce the overall, near-term
probability of a catastrophic event.

\section{Improved estimates}
\label{sectionimprovements} Our analysis places the 1968--2007
worldwide historic probability of a catastrophic event in the 11--35\%
range (see Table~\ref{tableestimates}) and none of the alternative or
covariate models provide any support for judging the size of 9/11 as
statistically unlikely. The wide confidence interval illustrates the
difficulty of obtaining precise estimates when accounting for model and
parameter uncertainty. That being said, our calculations could be
further refined to improve the overall estimates, incorporate
additional sources of uncertainty or address specific questions, by
relaxing portions of our i.i.d. treatment of event severities. We discuss
several such possibilities here, but leave their investigation for the future.

First, our algorithm assumes a stationary event generation process,
which is unlikely to be accurate in the long term. Technology,
population, culture and geopolitics are believed to exhibit
nonstationary dynamics and these likely play some role in event
severities. Thus, techniques for statistical forecasting in
nonstationary time series [\citet
{cairesferreira2005,clementshendry1999,shalizietal2010}] could be
used to identify subtle trends in the relevant covariates to make more
accurate forecasts.

Second, our algorithm is silent regarding efforts to prevent events or
mitigate their severity [\citet{kilcullen2010}]. However, the
historical impact of these processes is implicitly present in our
empirical data because only events that actually occurred were
recorded. Thus, our results may be interpreted as probabilities
conditioned on historical prevention or mitigation efforts. To the
extent that policies have an impact on incidence and severity, more
accurate estimates may be achievable by incorporating models of policy
consequences or interactions between different actors. Similarly, our
algorithm is silent regarding the actors responsible for events, and
incorporating models of organizational capabilities, proclivities,
etc. [\citet
{asalrethemeyer2008,clausetgleditsch2009,jacksonetal2005}] may
improve the estimates.

Finally, our approach is nonspatial and says little about where the
event might occur. Incorporating more fine-grained spatial structure,
for example, to make country-level or theatre-level estimates
[\citet
{zammit-mangionetal2012}] (as is now being done in seismology
[\citet
{leeetal2011}]), or incorporating tactical information, for example,
about specific CBRN attacks, may be possible. Such refinements will
likely require strong assumptions about many context-specific
factors [\citet{gartzke1999}], and it remains unclear whether accurate
estimates at these scales can be made. At the worldwide level of our
analysis, such contingencies appear to play a relatively small role in
the global pattern, perhaps because local-level processes are roughly
independent. This independence may allow large-scale general patterns
to emerge from small-scale contingent chaos [\citet
{lorenz1963,strogatz2001}] via a Central Limit Theorem averaging
process, just as regularities in birth rates exist in populations
despite high contingency for any particular conception. How far into
this chaos we can venture before losing general predictive power
remains unclear [\citet{rundleetal2011,wardetal2010}].

\section{Discussion} In many complex social systems, although large
events have outsized social significance, their rarity makes them
difficult to study. Gaining an understanding of such systems requires
determining if the same or different processes control the appearance
of small, common events versus large, rare events. A critical
scientific problem is estimating the true but unknown probability of
such large events, and deciding whether they should be classified as
statistical outliers. Accurate estimates can facilitate historical
analysis, model development and statistical forecasts.

The algorithm described here provides a principled and data-driven
solution for this problem that naturally incorporates several sources
of uncertainty. Conveniently, the method captures the tendency of
highly-skewed distributions to produce large events without reference
to particular generative mechanisms or strong assumptions about the
tail's structure. When properly applied, it provides an objective
estimate of the historical or future probability of a rare event, for
example, an event that has occurred exactly once.

Using this algorithm to test whether the size of the 9/11 terrorist
events, which were nearly six times larger than the next largest event,
could be an outlier, we estimated the historical probability of
observing at least one 9/11-sized event somewhere in the world over the
past 40 years to be 11--35\%, depending on the particular choice of
tail model used to fit the distribution's upper tail. These values are
much larger than any reasonable definition of a statistical anomaly and
thus the size of 9/11, which was nearly six times larger than the next
largest event, should not be considered statistically unlikely, given
the historical record of events of all sizes.

This conclusion is highly robust. Conditioning on the relative
frequency of important covariates [\citet{clausetetal2007}], such as
the degree of economic development in the target country, whether an
event is domestic or international, or the type of weapon used, we
recover similar estimates, with additional nuance. Large events are
probabilistically most likely to target economically developed nations,
be international in character and use explosives, arson, firearms or
unconventional weapons. Although chemical and biological events can
also be very large [\citet{cameron2000}], historically they are rare
at all sizes, and this outweighs the heaviness of their tail.

Furthermore, using only event data prior to 9/11 (as opposed to using
all available data sans 9/11), we find a similar overall historical
hazard rate. This suggests that the worldwide probability for large
events has not changed dramatically over the past few decades. In
considering three simple forecast scenarios for the next 10 years, we
find that the probability of another large event is comparable to its
historical level over the past 40 years. This risk seems unlikely to
decrease significantly without a large reduction in the number of
deadly terrorist events worldwide.

Of course, all such estimates are only as accurate as their underlying
assumptions, and our method treats event sizes as i.i.d. random variables
drawn from a stationary distribution. For complex social phenomena in
general, it would be foolish to believe this assumption holds in a very
strong sense, for example, at the micro-level or over extremely long
time scales, and deviations will lower the method's overall accuracy.
For instance, nonstationary processes may lower the global rate of
large events faster than smaller events, leading to overestimates in
the true probability of a large event. However, the i.i.d. assumption
appears to be statistically justified at the global spatial and
long-term temporal scales studied here. Identifying the causes of this
apparent i.i.d. behavior at the global scale is an interesting avenue for
future work.

The relatively high probability of a 9/11-sized event, both
historically and in the future, suggests that the global political and
social processes that generate large terrorist events may not be
fundamentally different from those that generate smaller, more common
events. Although the mechanism for event severities remains
unclear [\citet{clausetetal2010}], the field of possible explanations
should likely be narrowed to those that generate events of all sizes.

Independent of mechanistic questions, the global probability of another
large terrorist event remains uncomfortably high, a fact that can
inform our expectations [as with large natural disasters
\citet{gumbel1941,gutenbergrichter1944,reedmckelvey2002}] of how many
such events will occur over a long time horizon and how to
appropriately anticipate or respond to them. This perspective is
particularly relevant for terrorism, as classical models aimed at
predicting event incidence tend to dramatically underestimate event
severity [\citet{clausetetal2007}].

To conclude, the heavy-tailed patterns observed in the frequency of
severe terrorist events suggests that some aspects of this phenomenon,
and possibly of other complex social phenomena, are not nearly as
contingent or unpredictable as is often assumed. That is, there may be
global political and social processes that can be effectively described
without detailed reference to local conflict dynamics or the strategic
trade-offs among costs, benefits and preferences of individual actors.
Investigating these global patterns offers a complementary approach to
the traditional rational-actor framework [\citet{demesquita2003}] and
a new way to understand what regularities exist, why they exist, and
their implications for long-term stability.\looseness=1

\begin{appendix}

\section{Tail models}\label{appendixmethod}

The functional form and normalization of the tail model should follow
the type of empirical data used. For instance, if the empirical data
are real-valued, the power-law tail model has the form
%
\begin{equation}
\Pr(y | \alpha,\xmin) = \biggl({\alpha-1 \over\xmin} \biggr) \biggl(
{y \over\xmin} \biggr)^{-\alpha},\qquad \alpha>1, y\geq\xmin>0.
\end{equation}
Given a choice of $\xmin$, the maximum likelihood estimator for this
model is
%
\begin{equation}
\hat{\alpha} = 1 + n \Big/ \sum_{i=1}^{n}
\ln(x_{i}/\xmin).
\end{equation}
The severity of a terrorist attack, however, is given by an integer.
Thus, in our analysis of terrorist event severities, we use the
discrete form of the power-law distribution
%
\begin{equation}
\label{eqcontpl} \Pr(y | \alpha,\xmin) = y^{-\alpha} /
\zeta(\alpha,\xmin),\qquad \alpha>1, y\geq\xmin>0,
\end{equation}
where $\zeta(\hat{\alpha},\xmin)=\sum_{i=\xmin}^{\infty}
i^{-\alpha}$ is the generalized or incomplete zeta function. The MLE
for the discrete power law is less straightforward, being the solution
to the transcendental equation
%
\begin{equation}
\frac{\zeta'(\hat{\alpha},\xmin)}{\zeta(\hat{\alpha},\xmin)} = -\frac
{1}{n} \sum_{i=1}^{n}
x_{i}.
\end{equation}
However, it is straightforward to directly maximize the log-likelihood
function for the discrete power law in order to obtain $\hat{\alpha}$:
%
\begin{equation}
\mathcal{L}(\alpha) = -n \ln\zeta(\alpha,\xmin) - \alpha\sum
_{i=1}^{n} \ln x_{i}.
\end{equation}
Past work shows that the continuous model given by (\ref
{eqcontpl}) provides a reasonably good approximation to the discrete
case when $\xmin$ takes moderate values. In our own experiments with
this approximation, we find that when $\xmin\gtrsim10$ the difference
in estimated probabilities for observing one or more 9/11-sized events
between using the discrete versus continuous model is at most a few percent.

Estimates of $\xmin$ may be obtained using any of several existing
automatic methods [\citet
{danielssonetal2001,dekkersdehann1993,dreeskaufmann1998,breimanstonekooperberg1990,hancockjones2004}].
We use the Kolmogorov--Smirnov goodness-of-fit statistic minimization
(KS-minimization) technique [\citet
{clausetetal2007,clausetetal2009}]. This method falls in the
general class of distance minimization techniques for selecting the
size of the tail [\citet{reissthomas2007}], and was previously used
to analyze event severities in global terrorism. The KS
statistic [\citet{pressetal1992}]\vadjust{\goodbreak} is the maximum distance between the
CDFs of the data and the fitted model:
%
\begin{equation}
D = \max_{x\geq\xmin} \bigl\llvert S(x) - P(x) \bigr\rrvert,
\end{equation}
where $S(x)$ is the CDF of the data for the observations with value at
least $\xmin$, and $P(x)$ is the CDF of the maximum-likelihood
power-law model for the region $x\geq\xmin$. Our estimate $\hat
{x}_{\mathrm{min}}$ is then the value of $\xmin$ that minimizes $D$. In the
event of a tie between several choices for $\xmin$, we choose the
smaller value, which improves the statistical power of subsequent
analyses by choosing the larger effective sample size.

Our alternative tail models are the log-normal and the stretched
exponential distributions, modified to include a truncating parameter
$\xmin$. These distributions are normally defined on continuous
variables. The structure of their extreme upper tails for $\xmin=10$,
however, is close to that of their discrete versions, and the
continuous models are significantly easier to estimate from data. For
the results presented in the main text, we used the continuous
approximation of the upper tails for these models.

\section{Estimator accuracy}
\label{appendixaccuracy}

We quantify the expected accuracy of our estimates under two
experimental regimes in which the true probability of a catastrophic
event can be calculated analytically.
\begin{enumerate}
\item Draw $n$ values i.i.d. from a power-law distribution with $x_{\min
}=10$ and some~$\alpha$; define $x=\max_{i} \{ x_{i}\}$, the largest
value within that sample. This step ensures that we treat the synthetic
data exactly as we treated our empirical data and provides a
particularly challenging test, as the largest generated value exhibits
the greatest statistical fluctuations.
\item Draw $n-1$ i.i.d. values from a power-law distribution with $x_{\min
}=10$ and some~$\alpha$, and then add a single value of size $x$ whose
true probability of appearing under the generative model is $p=0.001$,
that is, we contaminate the data set with a genuine outlier.
\end{enumerate}

\begin{figure}

\includegraphics{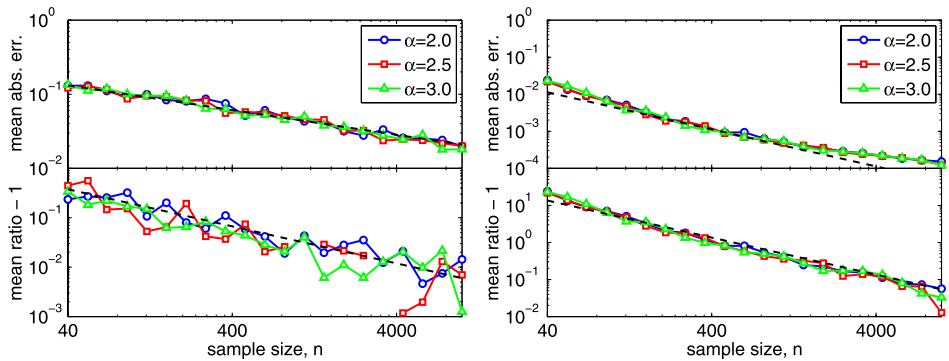}

\caption{The mean absolute error $\langle| \hat{p} - p | \rangle$
and mean relative error $\langle\hat{p}/p\rangle-1$ for \textup{(a)} $n$
values drawn i.i.d. from a stationary power-law distribution with $x_{\min
}=10$ and some $\alpha$, with the target size being the single largest
value in the draw, and for \textup{(b)} $n-1$ values to which we add a single
outlier (with true $p=0.001$). In both experiments, both types of
errors are small even for fairly small sample sizes and decay further
as $n$ increases.}
\label{figaccuracy}
\end{figure}

Figure \ref{figaccuracy} shows the results of both experiments, where
we measure the mean absolute error (MAE) and the mean ratio between
$\hat{p}$ and the true $p$. Even for samples as small as $n=40$
observations, the absolute error is fairly small and decreases with
increasing sample size $n$. In the first experiment, the error rate
decays like $O(n^{-1/3})$, approaching 0.01 error rates as $n$
approaches 5000 [Figure~\ref{figaccuracy}(a)], while in the second it
decays like $O(n^{-1})$ up to about $n=4000$, above which the rate of
decay attenuates slightly [Figure~\ref{figaccuracy}(b)].

%
%

Absolute deviations may mask dramatic relative errors, for example, if
the true probability is very close to one or zero (as in our
contaminated samples experiment). The mean ratio of $\hat{p}$ to $p$
would reveal such mistakes. The lower panels in Figure \ref
{figaccuracy} show that this is not the case: the estimation procedure
is close both in absolute and in relative terms. As the sample size
increases, the estimated probability converges on the true probability.
For contaminated data sets, the $\hat{p}/p$ can be fairly large when
$n$ is very small, but for sample sizes of a few hundred observations,
the method correctly estimates the relative size of the outlier's probability.

\section{Robustness checks}
\label{appendixrobustness}

We present three checks of the robustness of our probability estimates:
(i) using simple parametric models without the bootstrap, (ii) using an
alternative source of terrorist event data, and (iii) using event
covariates to refine the estimates. In each case, we find roughly
similar-sized estimates.

\subsection{Estimates using simple models}
\label{appendixsimple}

A simpler model for estimating the historical probability of a
9/11-sized or larger terrorist event assumes the following: (i) a
stationary generative process for event severities worldwide, (ii)
event sizes are i.i.d. random variables drawn from (iii) a power-law
distribution that (iv) spans the entire range of possible severities
($x_{\min}=1$), and (v) has a precisely-known parameter value $\alpha=2.4$.

A version of this model was used in a 2009 Department of
Defense-commis\-sioned JASON report on ``rare events'' [\citet
{mcmorrow2009}], which estimated the historical probability of a
catastrophic (9/11-sized or larger) terrorist event as 23\% over
1968--2006. The report used a slightly erroneous estimate of the power
law's normalizing constant, a slightly different estimate of $\alpha$
and a smaller value of $n$. Here, we repeat the JASON analysis, but
with more accurate input values.

Let $q(x)$ be the probability of observing a catastrophic event of size
$x$. With event severities being i.i.d. random variables drawn from a
fixed distribution $\Pr(y)$, the generation of catastrophic events can
be described by a continuous-time Poisson process with rate
$q(x)$ [\citet{boas2005}]. Approximating $x$ as a continuous variable,
in a sequence of $n$ such events, the probability $\hat{p}$ of
observing at least one of catastrophic severity is
%
\begin{eqnarray}
\label{eqpoisson}
\hat{p} & = & 1- \bigl[1-q(x) \bigr]^{n}
\nonumber\\[-8pt]\\[-8pt]
& \approx & 1-\mathrm{e}^{-n q(x)}.\nonumber
\end{eqnarray}

The rate $q(x)$ is simply the value of the complementary CDF at $x$,
and for a power-law distribution is given by
%
\begin{eqnarray}
\label{eqccdf} q(x) & = & \int_{x}^{\infty} \Pr(y)\,\d y
\nonumber
\\
& = & (\alpha-1)x_{\min}^{\alpha-1} \int_{x}^{\infty}
y^{-\alpha
} \,\d y
\\
& = & \biggl(\frac{x}{x_{\min}} \biggr)^{1-\alpha}
\nonumber
\end{eqnarray}
for $x\geq x_{\min}$. Substituting $x_{\min}=1$ and $\alpha=2.4$
yields the per-event probability of a catastrophic event $q(2749) =
0.0000153164$.

The RAND-MIPT database records $n=13274$ deadly events worldwide from
1968--2007; thus, substituting $n$ and $q(x)$ into (\ref
{eqpoisson}) yields a simple estimate of the probability of observing
at least one catastrophic event over the same time period $\hat{p} =
1-\mathrm{e}^{-13274 q(2749)} = 0.184$, or about 18\%.

However, this calculation underestimates the true probability of a
large event because the empirical distribution decays more slowly than
a power law with $\alpha=2.4$ at small values of $x$. Empirically
7.5\% of the 13,274 fatal events have at least 10 fatalities, but a
simple application of (\ref{eqccdf}) using $x=10$ shows that
our model predicts that only 4.0\% of events should be this severe.
Thus, events with $x\geq10$ occur empirically almost twice as often as
expected, which leads to a significant underestimate of $p$.

By restricting the power-law model to the tail of the distribution,
setting $x_{\min}=10$ and noting that only $n=994$ events had at least
this severity over the 40-year period, we can make a more accurate
estimate. Repeating the analysis above, we find $q(2749) =
0.0000288098$ and $\hat{p} = 0.318$, or about a 32\% chance of a
catastrophic event,\footnote{To make our reported per-event probabilities $q(x)$
consistent across models, we report them as $q(x)=\Pr(X\geq x |
X\geq x_{\min})\Pr(X\geq x_{\min})$, that is, the probability that a
tail event is catastrophic times the probability that the event is a
tail event. These values can be used with (\ref{eqpoisson})
to make rough estimates if the corresponding $n$ is the total number of
deadly events.} a value more in line with the estimates derived using our
bootstrap-based approach in the main text.

\subsection{Estimates using the Global Terrorism Database}
\label{appendixgtd}
An alternative source of global terrorism event data is the Global
Terrorism Database [\citet{gtd2011b}], which contains 98,112 events
worldwide from 1970--2007. Of these, 38,318 were deadly ($x>0$). Some
events have fractional severities due to having their total fatality
count divided evenly among multiple event records; we recombined each
group of fractional-severity events into a single event, yielding
38,255 deadly events over 38 years. Analyzing the GTD data thus
provides a check on our results for the RAND-MIPT data.

%
\begin{figure}

\includegraphics{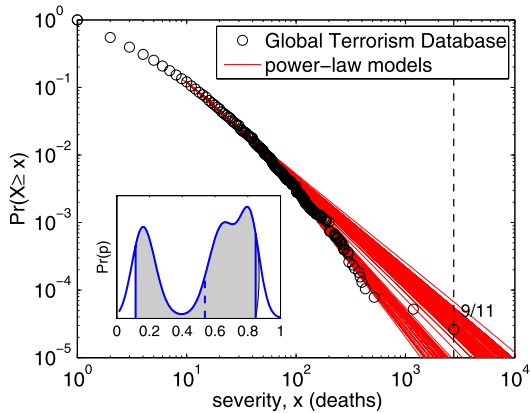}

\caption{Empirical distribution of event severities from the
GTD [START (\citeyear{gtd2011b})] with 100 power-law models, fitted to bootstraps
of the data. Inset shows the estimated distribution of binomial
probabilities $\Pr(\hat{p})$ for one or more catastrophic events.}
\label{figgtd}
\end{figure}
%

The largest event in the GTD is 9/11, with severity 2763, and the
second largest is the 13 April 1994 Rwandan massacre of Tutsi refugees,
with 1180 reported fatalities. This event is absent from the RAND-MIPT
data; its inclusion in the GTD highlights this data set's broader
definition of terrorism, which includes a number of genocidal or
criminal events.

The best fitting power-law model obtained using the methodology
of \citet{clausetetal2009} is $\hat{\alpha}=2.91\pm0.22$ and $\hat
{x}_{\min}=39$. The $p<0.1$ for this model may be attributable to the
unusually large number of perfectly round-number severities in the data
set, for example, 10, 20, 100, 200, etc., which indicates rounding
effects in the reporting. (These appear in Figure \ref{figgtd} as
small discontinuous drops in the complementary CDF at round-number
locations; true power-law distributed data have no preference for round
numbers and thus their presence is a statistically significant
deviation from the power-law form.)

Using the algorithm described in the main text with 10,000
bootstraps, we estimate a 38-year probability of at least one
catastrophic event as $\hat{p} = 0.534$ (with 90\% CI $[0.115,
0.848]$) or about a 53\% chance. Repeating our analysis using the two
alternative tail models yields only a modest decrease, as with the
RAND-MIPT data.

Figure \ref{figgtd} shows the empirical fatality distribution along
with 100 fitted power-law models, illustrating the heavy-tailed
structure of the GTD severity data. Notably, the maximum likelihood
estimate for $\alpha$ is larger here (indicating a less heavy tail)
than for the RAND-MIPT data. However, the marginal distribution $\Pr
(\hat{\alpha})$ is bimodal, with one mode centered on $\alpha=2.93$
and a second larger mode centered at roughly $\alpha=2.4$, in
agreement with the RAND-MIPT data. Furthermore, the failure of the
GTD-estimated $\hat{p}$ to be dramatically lower than the one
estimated using RAND-MIPT data supports our conclusion that the size of
9/11 was not statistically unlikely.

\subsection{Impact of event covariates}
\label{appendixcovariate}

\subsubsection{International versus domestic, and events prior to 1998}
\label{appendix1998}
Events in the RAND-MIPT database with dates before 1 January 1998 are
mainly international events, that is, the attacker's country of origin
differed from the target's country. Subsequent to this date, both
domestic and international events are included but their domestic
versus international character is not indicated. Analyzing events that
occurred before this breakpoint thus provides a natural robustness
check for our overall estimate: (i) we can characterize the effect that
domestic versus international events have on the overall estimate and
(ii) we can test whether the probability estimates have changed
significantly in the past decade.

\begin{figure}

\includegraphics{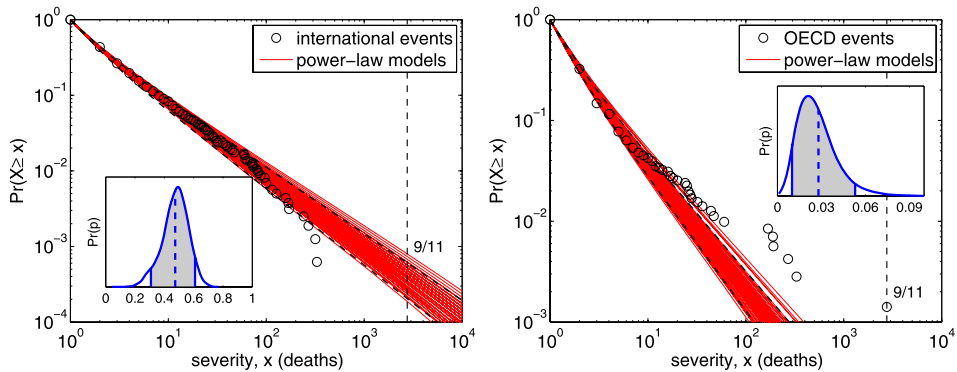}

\caption{Empirical distributions, with 100 power-law bootstrap models,
for \textup{(a)} international events (events from 1968--1997 in the RAND-MIPT
database) and \textup{(b)} events within the OECD nations; dashed
lines show the
90\% CI on $\hat{\alpha}$. Insets show the estimated distribution
$\Pr(\hat{p})$ with 90\% confidence intervals (shaded area) and
overall estimate (dashed line).}
\label{figcovariate}
\end{figure}

The pre-1998 events comprise 12\% of the RAND-MIPT database and exhibit
a more heavy-tailed distribution ($\hat{\alpha} = 1.92 \pm0.04$ and
$x_{\min} = 1$). Using 10,000 bootstraps, we estimate $\hat{p} =
0.475$ (90\% CI: $[0.309, 0.610]$) for at least one catastrophic
international event over the target period. Figure~\ref
{figcovariate}(a) shows the empirical distribution for international
events and the ensemble of fitted models, illustrating good visual
agreement with the empirical distribution.

%
%

The estimate for international-only data ($\hat{p} = 0.475$) is larger
than the estimate derived using the full data set ($\hat{p} = 0.347$),
although these values may not be as different as they seem, due to
their overlapping confidence intervals. Fundamentally, the larger
estimate is caused by the heavier-tailed distribution of the
international-only data.\vadjust{\goodbreak} Because the full data set includes these
international events, this result indicates that domestic events tend
to exhibit a lighter tail, and thus generate large terrorist events
with smaller probability. As a general guideline, subsets of the full
data set should be analyzed with caution, as their selection is
necessarily conditioned. The full data set provides the best estimate
of large events of all types.

\subsubsection{Economic development}
\label{appendixoecd}
A similar story emerges for deadly events in economically developed
nations, defined here as the member\vadjust{\goodbreak} countries of the Organisation for
Economic Co-operation and Development (OECD), as of the end of the
period covered by the RAND-MIPT data, which are 5.3\% of all deadly
events. The empirical distribution [Figure \ref{figcovariate}(b)] of
event severities shows unusual structure, with the upper tail ($x\geq
10$ fatalities) decaying more slowly than the lower tail. To handle
this oddity, we conduct two tests.

First, we consider the entire OECD data set, estimating both $\alpha$
and $x_{\min}$. Using 10,000 bootstraps yields $\hat{p}=0.028$ (with
90\% CI $[0.010, 0.053]$) or roughly a 3\% chance over the 40-year
period, which is slightly above our $p=0.01$ cutoff for a statistically
unlikely event. Figure \ref{figcovariate}(b) shows the resulting
ensemble of fitted models, illustrating that the algorithm is placing
very little weight on the upper tail. Second, we apply the algorithm
with a fixed $x_{\min}=10$ in order to focus explicitly on the
distribution's upper tail. In this case, 10,000 bootstraps yield $\hat
{p} = 0.225$, with 90\% CI as $[0.037, 0.499]$.

\subsubsection{Type of weapon}
\label{appendixweapons}
Finally, we consider the impact of the attack's weapon type, and we
generalize the estimation algorithm to the multi-covariate case. Events
are classified as (i) chemical or biological, (ii) explosives (includes
remotely detonated devices), (iii) fire, arson and firebombs, (iv)
firearms, (v) knives and other sharp objects, and (vi) other, unknown
or unconventional. Given the empirically observed distributions over
these covariates, we would like to know the probability of observing at
least one catastrophic-sized event from any weapon type.

This requires generalizing our Monte Carlo algorithm: let $(x,c)_{i}$
denote the severity $x$ and categorical covariate $c$ for the $i$th
event. Thus, denote the empirical data by $X=\{(x,c)_{i}\}$.
\begin{enumerate}
\item\label{step1b} Generate $Y$ by drawing $(y,c)_{j}$, $j=1,\ldots,n$,
uniformly at random, with replacement, from the original data $\{
(x,c)_{i}\}$ (sans the $\ell$ catastrophic events).
\item\label{step2b} For each covariate type $c$ in $Y$, jointly
estimate $\hat{x}_{\mathrm{min}}^{(c)}$ and the tail-model parameters
$\theta^{(c)}$, and compute $n_{\mathrm{tail}}^{(c)} = \#\{y_{j}\geq\hat
{x}_{\min}^{(c)}\}$.
\item\label{step3b} For each covariate type $c$ in $Y$, generate a
synthetic data set by drawing $n_{\mathrm{tail}}^{(c)}$ random deviates from
the fitted tail model with parameters $\hat{\theta}^{(c)}$.
\item If any of the covariate sequences of synthetic events includes at
least $\ell$ events of size $x$ or greater, set $\rho=1$; otherwise,
set it to zero.
\end{enumerate}
In applying this algorithm to our data, we choose $\ell=1$ and
$x=2749$, as with our other analyses. In step 2, we again use the
KS-minimization technique of \citet{clausetetal2009} to choose
$\xmin
$ and estimate $\theta$ for a power-law tail model via maximum
likelihood. Finally, as with the univariate version of the algorithm,
bootstrap confidence intervals may be obtained [\citet
{efrontibshirani1993}], both for the general hazard and the
covariate-specific hazard, by repeating steps 3 and 4 many times for
each bootstrap and tracking the distribution of binomial probabilities.

\begin{figure}

\includegraphics{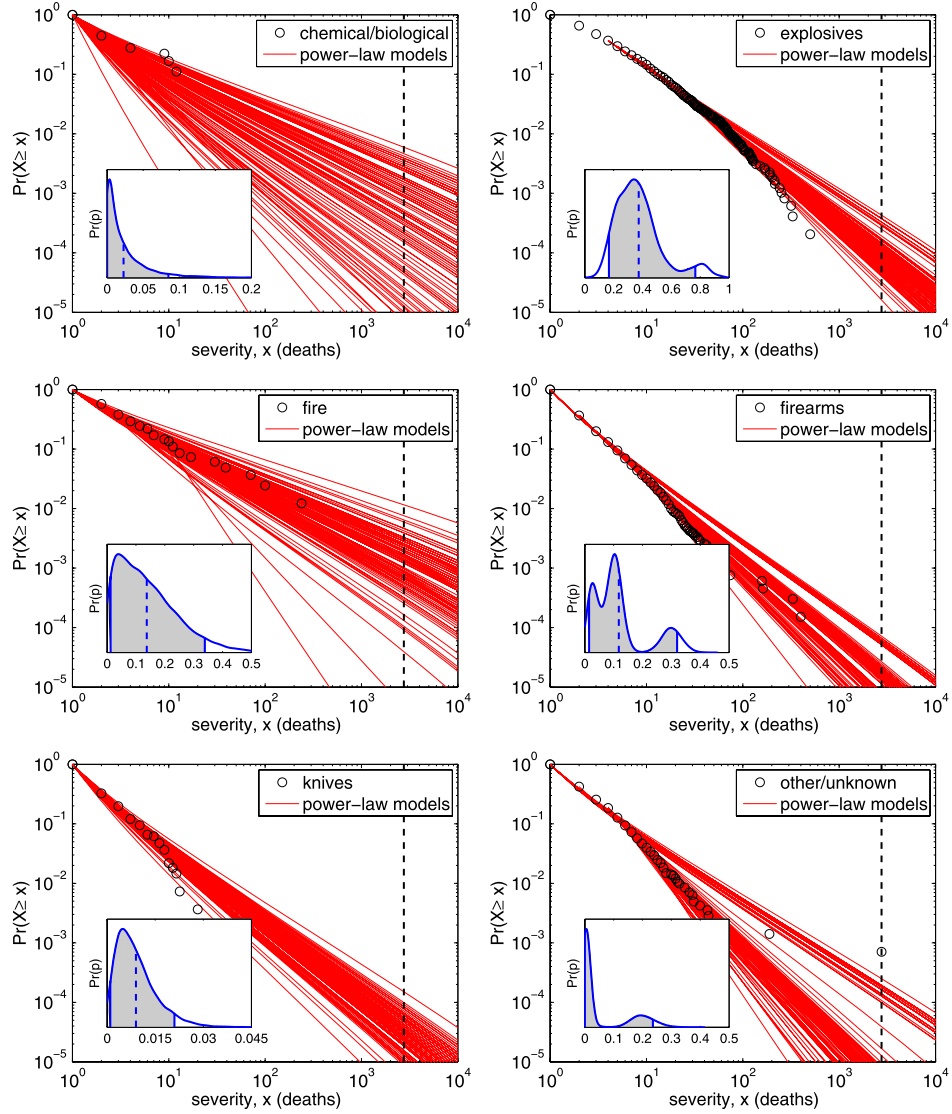}

\caption{Empirical distribution, with 100 power-law bootstrap models,
for events using \textup{(a)} chemical or biological, \textup{(b)} explosives (includes
remote detonation), \textup{(c)} fire, arson and firebombs, \textup{(d)} firearms, \textup{(e)}
knives or sharp objects, and \textup{(f)} other, unknown or unconventional
weapons. Insets: marginal distributions of estimated hazard rates $\Pr
(\hat{p})$, with the region of 90\% confidence shaded and the mean
value indicated by the dashed line.}
\label{figweapons}
\end{figure}

Using 10,000 bootstraps and drawing 1000 synthetic data sets from each
bootstrap, we estimate $\hat{p}=0.564$, with 90\% confidence intervals
of $[x,y]$. Again, this value is well above the cutoff for a 9/11-sized
attack being statistically unlikely. Figure \ref{figweapons}(a)--(f)
shows
the ensembles for each weapon-specific severity distribution. As a side
effect of this calculation, we may also calculate the probability that
a catastrophic event will be generated by a particular type of weapon.
The following table gives these marginal probability estimates, which
are greatest for explosives, fire, firearms and unconventional weapon types.

%
%

It is emphasized that these are historical estimates, based on the
relative frequencies of weapon covariates in the historical RAND-MIPT
data. If the future exhibits similar relative frequencies and total
number of attacks, then they may also be interpreted as future hazards,
but we urge strong caution in making these assumptions.

\begin{table}[h!]\vspace*{-6pt}
\tablewidth=195pt
\begin{tabular*}{\tablewidth}{@{\extracolsep{\fill}}lcc@{}}
\hline
\textbf{Weapon type} & \textbf{Historical} $\bolds{\hat{p}}$ & \textbf{90\%
CI} \\
\hline
Chem. or bio. & 0.023 & [0.000, 0.085] \\
Explosives & 0.374 & [0.167, 0.766] \\
Fire & 0.137 & [0.012, 0.339] \\
Firearms & 0.118 & [0.015, 0.320] \\
Knives & 0.009 & [0.001, 0.021] \\
Other or unknown & 0.055 & [0.000, 0.236]\\
\hline
Any & 0.564 & [0.338, 0.839]\\
\hline
\end{tabular*}\vspace*{-6pt}
\end{table}

(The sum of marginal probabilities exceeds that of the ``any'' column
because in some trials, catastrophic events are generated in multiple
categories.)
\end{appendix}

\section*{Acknowledgments}

The authors thank Cosma Shalizi, Nils Weidmann, Julian W\"ucherpfennig,
Kristian Skrede Gleditsch, Victor Asal, Didier Sornette, Lars-Erik
Cederman and Patrick Meier for helpful conversations.

Implementations of our numerical methods are available online at\break 
\url{http://www.santafe.edu/\textasciitilde aaronc/rareevents/}.


%

\printaddresses

\end{document}